\documentstyle[prl,aps]{revtex}
\input BoxedEPS.tex
\SetOzTeXEPSFSpecial
\HideDisplacementBoxes

\begin{document}

\twocolumn[\hsize\textwidth\columnwidth\hsize
           \csname @twocolumnfalse\endcsname
\title{Quantitative test of a quantum theory for the resistive 
transition in a superconducting single-walled carbon nanotube bundle} 
\author{Guo-meng Zhao$^{*}$} 
\address{Department of Physics and Astronomy, 
California State University, Los Angeles, CA 90032, USA}

\maketitle
\widetext

\maketitle
\widetext
\vspace{0.3cm}

\begin{abstract}

The phenomenon of superconductivity depends on the coherence of the 
phase of the superconducting order parameter. The resistive transition in 
quasi-one-dimensional (quasi-1D) superconductors 
is broad because of a large phase 
fluctuation. We show that the resistive transition of a superconducting  
single-walled carbon nanotube bundle is in quantitative agreement with 
the Langer-Ambegaokar-McCumber-Halperin (LAMH) theory. We also demonstrate that the  
resistive transition below $T^{*}_{c}$ = 0.89$T_{c0}$ is 
simply proportional to $\exp 
(-\frac{3\beta T^{*}_{c}}{T}(1-\frac{T}{T^{*}_{c}})^{3/2})$, where 
the barrier height has the same form as that predicted by the LAMH 
theory and $T_{c0}$ is the mean field superconducting transition 
temperature.

~\\
~\\
\end{abstract}
\narrowtext
]

The phenomenon of superconductivity depends on the coherence of the 
phase of the superconducting order parameter.  The phase coherence of 
the superconducting order parameter leads to the zero-resistance state.  
For three-dimensional (3D) bulk systems, the transition to the zero-resistance state occurs 
right below the mean-field superconducting transition temperature 
$T_{c0}$ such that the resistive transition is very sharp and the 
transition width is negligibly small.  In contrast,
the resistive transition in quasi-one-dimensional (quasi-1D) superconductors 
is broad because of a large superconducting 
fluctuation.  A quantum theory to describe the resistive transition in 
quasi-1D superconductors was developed by Langer, Ambegaokar, McCumber 
and Halperin (LAMH) \cite{Langer} over 30 years ago.  The theory is 
based on thermally activated phase slips (TAPS), which cause the 
resistance to decrease to zero exponentially. Experiments to test this 
theory were done by Lukens {\em et al.}\cite{Lukens} and Newbower {\em 
et al.} \cite{Newbower} on 
tin whiskers.  These are single-crystal, cylindrical specimens, 
typically $\sim$0.5 $\mu$m in diameter. The agreement bewteen the 
data and theory is satisfactory although the specimens are not truly
quasi-1D superconductors.  
Recently, great experimental efforts have been made to fabricate 
altrathin superconducting wires of amorphous MoGe, whose diameter can 
be smaller than 10 nm \cite{Tinkham1,Tinkham2}.  The resistive 
transitions of the altrathin wires appear to be in good agreement with the LAMH 
theory \cite{Tinkham2} although granularity and inhomogeneity may 
occur in these altrathin wires.

Now a question arises: Can we find a truly quasi-1D superconductor to 
quantitatively test the LAMH theory? Fortunately, it has been shown that the electronic structure of individual 
single-walled carbon nanotubes (SWNTs) has 1D nature.  The carbon nanotubes 
can be metallic or semiconducting, depending on their chiralities.  
The metallic individual carbon nanotubes should be ideal 1D superconductors if 
there are significant pairing interactions that overcome direct 
Coulombic interaction between conduction electrons.  Bundling these 
ideal 1D superconductors will lead to the formation of a quasi-1D 
superconducting wire with a much smaller superconducting fluctuation. Indeed, 
quasi-1D superconductivity below 0.5 K has been observed in a bundle of 
single-walled carbon nanotubes, which consists of about 350 tubes with mixed 
chiralities \cite{Kociak}. Here, we show that the resistive transition of this 
nanotube bundle is in quantitative agreement with the LAMH theory. We also demonstrate that the  
resistive transition below $T^{*}_{c}$ = 0.89$T_{c0}$ is 
simply proportional to $\exp 
(-\frac{3\beta T^{*}_{c}}{T}(1-\frac{T}{T^{*}_{c}})^{3/2})$, where 
the barrier height has the same form as that predicted by the LAMH 
theory.

In a theory developed by Langer, Ambegaokar,
McCumber and Halperin \cite{Langer}, phase slips occur via thermal 
activation, leading to a finite width for the resistive transition.  The 
resistance due to the TAPS is given by \cite{Book}
\begin{equation}
R_{TA} = \frac{h}{4e^{2}}\frac{\hbar\Omega}{k_{B}T}\exp [-\Delta 
F_{\circ}(T)/k_{B}T],
\end{equation}
where the attempt frequency $\Omega$  is  \cite{Langer}
\begin{equation}
\Omega = \frac{\sqrt{3}}{2\pi^{3/2}}\frac{L}{\xi} \sqrt{\frac{\Delta 
F_{\circ}(T)}{k_{B}T}}\frac{1}{\tau_{GL}}.
\end{equation}
Here $L$ is the length of the wire, $\xi (T)$ is the coherence length, and 
$\hbar/\tau_{GL} = (8/\pi) k_{B}(T_{c0} -T)$.  The barrier energy $\Delta 
F_{\circ}(T)$ is
\begin{equation}
\Delta F_{\circ}(T) = \frac{8\sqrt{2}}{3}\frac{H_{c}^{2}(T)}{8\pi}A\xi,
\end{equation}
where $H_{c}^{2}(T)/8\pi$ is the condensation energy, $A$ is the 
cross-section area of the wire, and the critical field near 
$T_{c0}$ is given by $H_{c}(T) = 1.73H_{c}(0)(1-T/T_{c0})$ within the 
BCS theory \cite{Book}.  Using $\xi (T) = \xi (0)(1-T/T_{c0})^{-1/2}$ 
(Ref.~\cite{Book}), we then have
\begin{equation}
\frac{\Delta F_{\circ}(T)}{k_{B}T} = 
\frac{3cT_{c0}}{T}(1-\frac{T}{T_{c0}})^{3/2},
\end{equation}
where 
\begin{equation}
c=\frac{\Delta F_{\circ}(0)}{k_{B}T_{c0}}
=\frac{8\sqrt{2}}{3}\frac{H_{c}^{2}(0)}{8\pi k_{B}T_{c0}}A\xi 
(0).
\end{equation}
Combining the above equations, we finally get 
\begin{equation}\label{Er}
R_{TA} = \frac{m}{T^{1.5}}(1-\frac{T}{T_{c0}})^{9/4}\exp 
[-\frac{3cT_{c0}}{T}(1-\frac{T}{T_{c0}})^{3/2}],
\end{equation}
with
\begin{equation}\label{Em}
m = 2.55T_{c0}(3cT_{c0})^{1/2}\frac{L}{\xi (0)}, 
\end{equation}
where $m$ is in the unit of k$\Omega$K$^{3/2}$. We would like to mention 
that the exponent in Eq.~\ref{Er} is factor of 3 larger than that in a 
similar formula deduced in Ref.~\cite{Tinkham2}. It is possible that 
the authors in Ref.~\cite{Tinkham2} missed the prefactor of 1.73 in 
$H_{c}(T)$.

The condensation energy at zero temperature $H_{c}^{2}(0)/8\pi$ is equal to 
$N(0)\Delta^{2}(0)/2$ within the BCS theory, where $N(0)$ is the density 
of states near the Fermi level and $\Delta (0)$ is the 
superconducting gap at zero temperature. For a single metallic SWNT with 
two transverse channels, $N(0)A= 4/3\pi 
a_{C-C}\gamma_{\circ}$ (Ref.~\cite{RochePRB}), $\hbar v_{F} = 
1.5a_{C-C}\gamma_{\circ}$ (Ref.~\cite{Mintmire}), where 
$\gamma_{\circ}$ is the hopping integral and $a_{C-C}$ (0.142 nm) is the bonding 
length.  Using the BCS relations: $\xi_{BCS} = \hbar v_{F}/\pi\Delta 
(0)$ and 
$\Delta (0)/k_{B}T_{c0}$ =1.76, and the above relations, one can readily show 
that
\begin{equation}
c = 0.68\frac{\xi (0)}{\xi_{BCS}}. 
\end{equation}

If a bundle of single-walled nanotubes or a multi-walled nanotube consists 
of $N_{ch}$ transverse channels, then 
\begin{equation}\label{Ec}
c = 0.34N_{ch}\frac{\xi (0)}{\xi_{BCS}}.
\end{equation}

For two-probe or four-probe measurements on carbon nanotubes with finite 
transverse channels, the total resistance is $R(T) = R_{0}+R_{tube}$, 
where $R_{tube}$ is  the on-tube 
resistance and $R_{0}$ = $R_{t}$ = $R_{Q}/tN_{ch}$ (tunneling resistance) for four-probe 
measurements, or $R_{0}$ = $R_{Q}/tN_{ch}+R_{c}$ for two probe 
measurements \cite{Sheo}. Here $t$ is the transmission 
coefficient ($t \leq$ 1), $R_{Q}$ = $h/2e^{2}$ = 12.9 k$\Omega$ is 
the resistance quantum, and $R_{c}$ is the contact resistance.  Both $R_{c}$ and $R_{t}$ should 
be temperature independent.  For ideal contacts, $R_{c}$ = 0 and 
$t$ = 1,  so $R_{t}$ = 12.9 
k$\Omega$/$N_{ch}$ for a bundle comprising $N_{ch}$ transverse channels. 
For quasi-1D systems,  $N_{ch}$ is always finite such that  $R(T)$ 
never goes to zero even if the on-tube resistance is zero. Only 
if $N_{ch}$ goes to infinity, as in the bulk 3D systems, $R_{t}$ becomes zero such that four 
probe resistance can go to zero below the superconducting transition temperature.

Fig.~1 shows the two-probe resistance data for a SWNT bundle that consists of about 
350 tubes \cite{Kociak}.  One can see that the resistance starts to drop below about 
0.5 K, decreases more rapidly below  $T_{c0}$ $\simeq$ 0.44 K and saturates to a value 
of 74 $\Omega$. From the saturated value of $R_{0}$ = 74 $\Omega$, and the 
relation: $R_{0}$= $R_{Q}/tN_{ch}+R_{c}$, one can 
easily find that more than 174 transverse channels are connected to 
the electrodes and participate in electrical transport. This implies that more than 87 metallic-chirality 
superconducting SWNTs take part in electrical transport. Considering 
the fact that one third of tubes should have metallic chiralities and 
become superconducting, we find the total number ($N_{m}$) of the 
superconducting tubes to be 117, implying that $t$ $\geq$ 0.74. The 
value of $N_{m}$ can be also deduced from the measured current 
$I_{c}^{*}$ at which the last resistance jump occurs. The $I_{c}^{*}$ corresponds to 
the critical current for a superconducting wire without disorder and 
with the same number of the transverse channels \cite{Kociak}. For 
metallic chirality superconducting carbon nanotubes, one can easily 
deduce that \cite{Zhao} $I_{c}^{*} = 
7.04k_{B}T_{c0}N_{m}/eR_{Q}$. With $I_{c}^{*}$ = 2.4 $\mu$A 
(Ref.~\cite{Kociak}) and  $T_{c0}$ = 
0.44 K, we have $N_{m}$ = 116, in remarkably good agreement with the 
value deduced above. 

\begin{figure}[htb]
\ForceWidth{7cm}
	\centerline{\BoxedEPSF{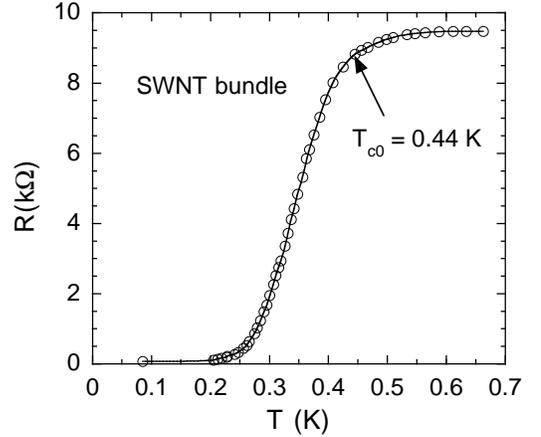}}
	\vspace{0.4cm}
\caption[~]{The temperature dependence of the two-probe resistance for a SWNT bundle that consists of about 
350 tubes.  The data are extracted from Ref.~\cite{Kociak}.}
\end{figure}

In Fig.~2, we fit the resistance data below 
0.88$T_{c0}$ by 
\begin{equation}\label{Ee} 
R = 74+\alpha\exp 
[-\frac{3\beta T^{*}_{c}}{T}(1-\frac{T}{T^{*}_{c}})^{3/2}]. 
\end{equation}
Here the first term is the sum of the tunneling and contact 
resistances discussed above, and the second term is the on-tube 
resistance which has a similar exponential dependence on $T$ as Eq.~\ref{Er} but with 
a temperature independent prefactor. We can see that the fit is excellent with the fitting parameter 
$\beta$ = 2.99$\pm$0.05 and $T^{*}_{c}$ = 0.394$\pm$0.002 K. Reducing or increasing the temperature region for the fit tends to 
worsen the fit quality. Therefore, the on-tube resistance 
goes to zero exponentially below $T^{*}_{c}$ = 0.89$T_{c0}$. The 
microscopic origin of this simple exponential dependence up to 0.89$T_{c0}$ 
is not clear, so we consider Eq.~\ref{Ee} only as an empirical formula. 

\begin{figure}[htb]
\ForceWidth{7cm}
	\centerline{\BoxedEPSF{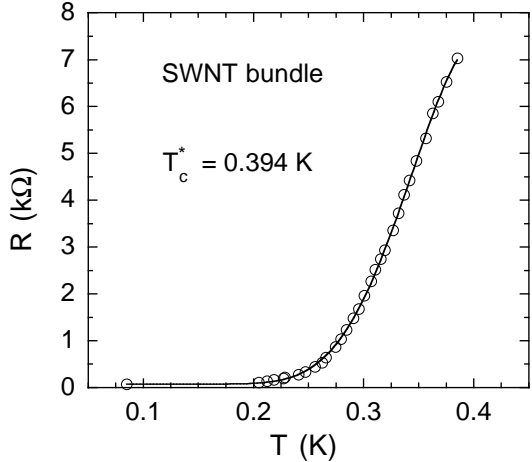}}
	\vspace{0.4cm}
\caption[~]{The temperature dependence of the two-probe resistance for 
a SWNT bundle below 0.88$T_{c0}$. 
The solid line is the curve best fitted by Eq.~\ref{Ee} 
with $\beta$ = 2.99$\pm$0.05 and $T^{*}_{c}$ = 0.394$\pm$0.002 K. 
Reducing or increasing the temperature region for the fit tends to 
worsen the fit quality. It is striking that the 
on-tube resistance below 0.88$T_{c0}$ decreases exponentially to 
zero.}
\end{figure}

We also try to fit the 
resistance below 0.88$T_{c0}$ by 
\begin{equation}\label{MH} 
R = 74+\frac{m}{T^{1.5}}(1-\frac{T}{T_{c0}})^{9/4}\exp 
[-\frac{3cT_{c0}}{T}(1-\frac{T}{T_{c0}})^{3/2}].
\end{equation}
Here the second term is the on-tube resistance which is the same as 
Eq.~\ref{Er} predicted by the LAMH theory. We find that the fit is not good and the fitting parameters have no 
quantitative agreement with the LAMH theory. This is because the LAMH 
theory is only applied to the temperature region where the barrier 
height is far larger than $k_{B}T$ so that current carrying states 
involved are truly metastable \cite{MMP}. The estimated region of validity 
for the  LAMH theory is below 0.07$R_{N}$ for dirty wires 
where the mean free path $l$ $<$$<$ $\xi_{BCS}$ (Ref.~\cite{MMP}). For the SWNT bundle, the 
condition of $l$ $<$$<$ $\xi_{BCS}$ is well satisfied, as seen below. In Fig.~3, we fit the resistance data 
below 0.06$R_{N}$ by Eq.~\ref{MH}. Here we have ignored the normal 
conduction because $R_{TA}$ is a factor of about 20 smaller than 
$R_{N}$. One can see that the 
fitting is very good with the fitting parameters: $m$ = 26.6$\pm$4.7 k$\Omega 
K^{1.5}$ and $c$ = 3.08$\pm$0.13. It is remarkable that the value of 
$c$ is nearly the same as the value of $\beta$ (2.99$\pm$0.05) deduced 
above from a simple exponential fit (Eq.~\ref{Ee}).  We will see below that the fitting 
parameters $c$ and $m$ are in quantitative agreement with the  LAMH theory.

\begin{figure}[htb]
\ForceWidth{7cm}
	\centerline{\BoxedEPSF{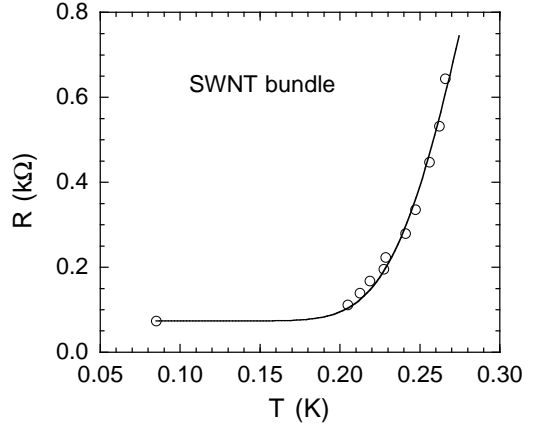}}
	\vspace{0.4cm}
\caption[~]{The temperature dependence of the two-probe resistance for a SWNT 
bundle below 0.06$R_{N}$. 
The solid line is the curve fitted to the data below 0.06$R_{N}$ by 
Eq.~\ref{MH} predicted by the LAMH theory. The estimated region of validity 
for the  LAMH theory is below 0.07$R_{N}$ for dirty wires where 
$l$ $<$$<$ $\xi_{BCS}$ (Ref.~\cite{MMP}). The condition of $l$ $<$$<$ $\xi_{BCS}$ is well satisfied in the SWNT bundle (see text). }
\end{figure}

From the values of $m$, $c$, and $T_{c0}$, we can evaluate the 
zero-temperature coherence length $\xi (0)$ using Eq.~\ref{Em}. 
Substituting $m$ = 26.6 k$\Omega 
K^{1.5}$, $c$ = 3.08, $T_{c0}$ = 0.44 K, and $L$ = 10000 \AA~ into 
Eq.~\ref{Em}, we obtain $\xi (0)$ = 850 \AA. From the measured 
$R_{N}(0.25K)/L$ = 12 k$\Omega$/$\mu$m (Ref.~\cite{Kociak}) and the relation 
$R_{N}/L$ = $R_{Q}/2N_{m}l$ (Ref.~\cite{Bend}), we can calculate the mean free path 
$l$ at 0.25 K. With $N_{m}$ = 117, we get $l$ = 46 \AA. Substituting $l$ = 46 
\AA~ and $\xi (0)$ = 850 \AA~ into the dirty-limit 
formula \cite{MMP}: $\xi (0) = 0.85\sqrt{\xi_{BCS}l}$, we have $\xi_{BCS}$ = 21739 
\AA. With $\xi_{BCS}$ = 21739 \AA, $\xi (0)$ = 850 \AA, and $N_{ch}$ = 
2$N_{m}$ = 334, we calculate $c$ = 3.11 from Eq.~\ref{Ec}. It is remarkable that the 
calculated value of $c$ from Eq.~\ref{Ec} is in quantitative agreement 
with the value (3.08) deduced from the fitting.

We can also estimate the Fermi velocity $v_{F}$ from the deduced value 
of $\xi_{BCS}$ and the formula $\xi_{BCS} = 0.18\hbar v_{F}/k_{B}T_{c0}$. 
With $\xi_{BCS}$ = 21739 
\AA~ and $T_{c0}$ = 0.44 K, we get  $\hbar v_{F}$ = 4.6 eV\AA. Then 
we estimate $\gamma_{\circ}$ = 2.16 eV from $\hbar v_{F} = 
1.5a_{C-C}\gamma_{\circ}$. This value is very close to an 
independent estimate (2.26 eV) from the measured semiconducting gap for 
a $d$ = 1.34 nm SWNT \cite{Collins}.

The deduced value of $\xi (0)$ = 850 \AA~ is also in excellent 
agreement with the measured critical current $I_{c}$ for this SWNT 
bundle. For a diffusive superconducting wire, the critical current 
$I_{c}$ is given by \cite{Egg}
\begin{equation}\label{Ei}
I_{c}= \frac{\Delta (0)}{eR_{N}}\frac{L}{\xi (0)}.
\end{equation}
Using $\Delta (0) = 1.76k_{B}T_{c0}$ and substituting $R_{N}(0.1K)/L$ =12.5 
k$\Omega$/$\mu$m (Ref.~\cite{Kociak}) and $\xi (0)$ = 850 \AA~ into Eq.~\ref{Ei}, we find 
$I_{c}$ = 62.4 nA, which is very close to the measured value (62 nA) at 
0.1 K (Ref.~\cite{Kociak}).

In summary, we have shown that the observed resistive transition of a superconducting  
carbon nanotube bundle is in quantitative agreement with 
the LAMH theory. We have also 
demonstrated that the  
resistive transition below $T^{*}_{c}$ = 0.90$T_{c0}$ is 
simply proportional to $\exp 
(-\frac{3\beta T^{*}_{c}}{T}(1-\frac{T}{T^{*}_{c}})^{3/2})$, where 
the barrier height has the same form as that predicted by the LAMH 
theory.

~\\ 
 ~\\
* Correspondence should be addressed to gzhao2@calstatela.edu.


\begin{thebibliography}{99}

\bibliographystyle{prsty}
\bibitem{Langer}J.  S.  Langer and V.  Ambegaokar, 
Phys.  Rev.  {\bf 164}, 498 (1967); D.  E.  McCumber and B.  I.  
Halperin, Phys.  Rev.  B {\bf 1}, 1054 (1970).
\bibitem{Lukens}J. E. Lukens, R. J. Warburton, and W. W. Webb, Phys. Rev. Lett. 
{\bf 25}, 1180 (1970).
\bibitem{Newbower}R. S. Newbower, M. R. Beasley, and M. Tinkham, Phys. 
Rev. B {\bf 5}, 864 (1972).
\bibitem{Tinkham1}A.  Bezryadin, C.  N.  Lau, and M.  
Tinkham, Nature (London) {\bf 404}, 971 (2000).
\bibitem{Tinkham2}C. N. Lau, N. Markovic, M. Bockrath, A. Bezryadin, and 
M. Tinkham, Phys. Rev. Lett. {\bf 87}, 217003 (2001).
\bibitem{Kociak} M. Kociak, A.Yu. Kasumov, S. Gueron, B. Reulet, I. I. 
Khodos, Yu. B. Gorbatov, V. T. Volkov, L. Vaccarini, and H. Bouchiat, Phys. 
Rev. Lett. {\bf 86}, 2416 (2001). 
\bibitem{Book} M.~Tinkham, Introduction to Superconductivity (McGraw-Hill, 1996).
\bibitem{RochePRB}Stephan Roche and Riichiro Saito, Phys.  Rev.  B {\bf 
59}, 5242 (1999).
\bibitem{Mintmire}J. W. Mintmire and C. T. White, Phys.  Rev.  Lett.  
{\bf 81}, 2506 (1998).
\bibitem{Sheo}C.  
Sch\"onenberger, A.  Bachtold, C.  Strunk, J.-P.  Salvetat, L.  Forro, 
Appl.  Phys.  A {bf 69}, 283 (1999). 
\bibitem{Zhao} G. M. Zhao, cond-mat/0208198.
\bibitem{MMP}J. R. Trucker and B. I. Halperin, Phys. Rev. B {\bf 3}, 
3768 (1971).  
\bibitem{Bend}L. X. Bendedict, V. H. Crespi, S. G. Louie, and M. L. 
Cohen, Phys. 
Rev. B {\bf 52}, 14935 (1995).

\bibitem{Collins}P. G. Collins, M. S. Arnold, P.  Avouris {\bf 292}, 
706 (2001).


\bibitem{Egg}M. Ferrier, A. De Martino, A. Kasumov,S. Gueron, M. Kociak, R. Egger, and 
H. Bouchiat, cond-mat/0405449.

\end{thebibliography}
\end{document}